\documentclass[11pt]{article} 

\usepackage[utf8]{inputenc} 

\usepackage{geometry} 
\usepackage{graphicx} 

\usepackage{booktabs} 
\usepackage{array} 
\usepackage{paralist} 
\usepackage{verbatim} 
\usepackage{subfig} 
\usepackage{amsmath}
\usepackage{amsthm}

\usepackage{fancyhdr} 
\usepackage{sectsty}
\allsectionsfont{\sffamily\mdseries\upshape} 

\usepackage[nottoc,notlof,notlot]{tocbibind} 
\usepackage[titles,subfigure]{tocloft} 

\newcommand{\argmax}[1]{\underset{#1}{\operatorname{arg}\,\operatorname{max}}\;}

\newtheorem*{theorem}{Theorem}

\title{Fat Tailed Factors}
\author{Jan Rosenzweig\footnote{Pine Tree Funds, 107 Cheapside, London EC2V 6DN, j.rosenzweig@pinetree-funds.com; King's College London, Strand, London WC2R 2LS, jan.rosenzweig@kcl.ac.uk}}
\date{}

\begin{document}

\maketitle

\abstract{Standard, PCA-based factor analysis suffers from a number of well known problems due to the random nature of pairwise correlations of asset returns. We analyse an alternative based on ICA, where factors are identified based on their non-Gaussianity, instead of their variance. \\
Generalizations of portfolio construction to the ICA framework leads to two semi-optimal portfolio construction methods: a {\it fat-tailed portfolio}, which maximises return per unit of non-Gaussianity, and the {\it hybrid portfolio}, which asymptotically reduces variance and non-Gaussianity in parallel.\\
For fat-tailed portfolios, the portfolio weights scale like performance to the power of $1/3$, as opposed to linear scaling of Kelly portfolios; such portfolio construction significantly reduces portfolio concentration, and the winner-takes-all problem inherent in Kelly portfolios.\\
For hybrid portfolios, the variance is diversified at the same rate as Kelly PCA-based portfolios, but  excess kurtosis is diversified much faster than in Kelly,  at the rate of $n^{-2}$ compared to Kelly portfolios' $n^{-1}$ for increasing number of components $n$.\\}


{\bf Key messages:}
\begin{itemize}
\item tail-risk-based portfolios can address a number of know problems of covariance-based portfolios
\item when portfolios are optimized for tail risk, weights of components scale sub-linearly with their performance, reducing portfolio concentration and the winner-takes-all problem
\item portfolios optimized for tail risk manage variance as efficiently as covariance-based portfolios.
\end{itemize}

Word count: 4,437\\

Number of figures: 4\\

Number of tables: 0\\

\section{Introduction}

Analysis of portfolio returns is usually performed with reference to the relationship between the portfolio's returns vector and its covariance matrix, a practice stemming back from Markowitz's Modern Portfolio Theory in the 1950s \cite{mark1, mark2}. Briefly, the classical theory follows the mean-variance approach, whereby orthogonal assets are allocated in inverse proportion to their volatility, multiplied by their Sharpe ratio \cite{roll1}. 

Most assets are, of course, not always orthogonal to each other. The approach above is then applied to a suitable orthogonalization of the asset universe, which is invariably done using the Principal Component Analysis (PCA) \cite{pca}. Fast algorithms for extraction of principal components are routinely used to partially invert covariance matrices and construct approximate optimal portfolios.

Covariance based methods such as the PCA, however, suffer from a number of well-known problems. The main problem is that correlations of asset returns are not particularly robust, and they may even arise randomly \cite{avellaneda}. There have been a number of approaches aimed at alleviating that problem (see \cite{avellaneda, principal, shkolnik} and references therein). Briefly, only the principal eigenportfolio, corresponding to the first principal component (PC) is sufficiently stable and reliable to have a meaningful economic interpretation \cite{principal}, and higher PCs are generally neither stable, or interpretable.

We take a somewhat different approach, based on the Independent Component Analysis (ICA) \cite{icawiki}. In signal processing, ICA is a computational method for separating a multivariate signal into additive subcomponents. This is done by assuming that the subcomponents are non-Gaussian signals and that they are statistically independent from each other. ICA is a special case of blind source separation. A common example application is the "cocktail party problem" of listening in on one person's speech in a noisy room \cite{hyva}.

ICA is a robust method widely used in engineering signal processing applications ranging from  acoustic signal processing \cite{audio}, visual processing \cite{visual} and telecommunications \cite{telecom}, to electroencephalography \cite{eeg} and many others. 

Practically, ICA extracts factors based on their non-Gaussianity, rather than on their variance, as PCA. Traditionally, the chosen measure of non-Gaussianity was excess kurtosis. Other measures of non-Gaussianity have also been popular, such as various empirical approximations to negentropy \cite{fastica}. 

In enginering applications ranging from acoustic signal processing \cite{audio}, visual processing \cite{visual}, telecomunications \cite{telecom}, electroencephalography \cite{eeg} and many others, ICA is generally considered to be a considerably more powerful tool than PCA, and it is a {\it de facto} standard in a wide range of applications. 

The strength of the ICA approach used here is that, unlike PCA, ICA makes no assumptions about the underlying universe following a multivariate Gaussian distribution \cite{choi}; it therefore does not break down when the underlying data sufficiently deviates from Gaussianity. 

We show that the ICA arises naturally when trying to optimize a portfolio against a penalty proportional to its kurtosis, as opposed to its variance in the classical case; the resulting optimal portfolio has weights of independent components (IC)  proportional to the cubed root of the ratio of the mean to the kurtosis of each component. The scaling of the portfolio weight with its mean return is thus sub-linear, as opposed to linear in the classical case.

This paper is set out as follows. Section 2 outlines the the main differences between PCA and ICA, and sets out the notation. Section 3 is dedicated to {\it Fat-Tailed Portfolios}, a generalization of Kelly portfolios to the non-Gaussianity framework. Section 4 analyses {\it Hybrid Portfolios}, a class of kurtosis- and variance- minimizing portfolios. Section 5 looks at a practical analysis of S\&P500 constituents over a 12 year period, and Section 6 is the Discussion.

\section{PCA vs ICA}

Let $S^{(1)}_{t}..S^{(N)}_{t}$ denote prices of $N$ assets  at time $t$, forming a vector of asset prices ${\bf S}_{t}$, with a corresponding vector of returns $d{\bf S}_{t}$. The returns  $d{\bf S}_{t}$ are usually taken as lognormal returns ($d{\bf S}_{t} = \ln \left( {\bf S}_{t+\Delta t} / {\bf S}_{t} \right)$), but all the analysis below equally holds for  normal returns ($d{\bf S}_{t} =   {\bf S}_{t+\Delta t} / {\bf S}_{t} - 1 $), or price returns  ($d{\bf S}_{t} =   {\bf S}_{t+\Delta t} - {\bf S}_{t} $).

Let ${\bf w}^{(i)}$ be portfolio weight $N$-vectors, and  ${\Pi}^{(i)}$ be the resulting portfolios,
\begin{equation}
\Pi_{t}^{(i)} = {\bf w}^{(i)} .  {\bf S}_{t} \label{portcomp}
\end{equation}
We denote the variances and kurtoses of the returns of the $i$'th portfolio $\Pi_{t}^{(i)}$ by $\sigma^{(i)}$  $\kappa^{(i)}$, respectively. 

Both {\it Principal Component Decomposition} (PCA) and {\it Independent Component Decomposition} (ICA) are decompositions of the form (\ref{portcomp}), and they can both be obtained iteratively.

For  the $i$th component of the PCA, the unit weight vector  ${\bf w}^{(i)}$ is selected so as to maximise the variance $\sigma^{(i)}$,   the residues are projected to the hyperplane orthogonal to the resulting component $\Pi^{(i)}$, and the iteration then proceeds to the component $i+1$ for the residues.

Note that the resulting principal components are generally long-short, and that we make no assumptions on positivity of weights, or them adding up to a positive number. They are only normalized to unit length.

For the $i$th component of the ICA, the process is analogous, except that the  weight vector  ${\bf w}^{(i)}$ is selected so as to maximise the kurtosis $\kappa^{(i)}$ instead of the variance $\sigma^{(i)}$. Also, by convention, the weight vectors are normalised differently in PCA and ICA. While PCA weight vectors are normalised to the same portfolio weight (typically unity), ICA weights are normalised to the same variance (typically the variance of the first component, whose weight vector is normalised to unity). 
 
In further text, we denote the $i$th principal component as $PC^{(i)}$, and the $i$th independent component as $IC^{(i)}$. The respective means, volatilities and kurtoses of the returns of the  principal and independent components are denoted $\mu_{PCA}^{(i)}$, $\mu_{ICA}^{(i)}$, $\sigma_{PCA}^{(i)}$, $\sigma_{ICA}^{(i)}$, $\kappa_{PCA}^{(i)}$ and $\kappa_{ICA}^{(i)}$, respectively.

\section{Fat Tailed Portfolios}

We denote by
$${\bf m} = E\left(d{\bf S}\right),$$
$${\bf V} = E\left( d{\bf S} \otimes d{\bf S}\right) - {\bf m}^{2}$$
$${\bf K} = E\left( (d{\bf S} - {\bf m}) \otimes (d{\bf S} - {\bf m})\otimes (d{\bf S} - {\bf m})\otimes (d{\bf S} - {\bf m})\right)  - 3 {\bf V} \otimes {\bf V}$$
the return, covariance and excess cokurtosis of the joint distribution of the returns of the asset process ${\bf S}$.

The Kelly criterion for portfolio construction is easily obtained  by looking for the weights that maximize the portfolio return, while penalising for portfolio variance:
\begin{equation}
{\bf w} = \argmax{\bf w} \left( {\bf w}.{\bf m} - \lambda\ {\bf w}.{\bf V}.{\bf w} \right) \label{kellymin}
\end{equation}
for some risk aversion parameter $\lambda$.

Differentiating the right hand side of (\ref{kellymin}) wrt ${\bf w}$ and setting the result to zero yieds
the familiar Kelly criterion 
\begin{equation}
{\bf w} \propto {\bf V}^{-1}.{\bf m} \label{kelly}
\end{equation}
where the constant of proportionaity depends on the investor's risk aversion. Equation (\ref{kelly}) is easily solved using  PCA; noting that each eigenvector of ${\bf V}$ corresponds to the weights vector of a principal component, and that the corresponding eigenvalue corresponds to its variance, we can write the optimal portfoio weights in terms of PCs as
\begin{equation}
 w_{PCA}^{(i)} \propto \frac{ \mu_{PCA}^{(i)}}{{\sigma_{PCA}^{(i)}}^{2}}  \label{kellyPCA}
\end{equation}
The optimal portfolio is given as
\begin{equation}
 \Pi \propto \sum_{i} w_{PCA}^{(i)} PC^{(i)} = \sum_{i} \frac{ \mu_{PCA}^{(i)}}{{\sigma_{PCA}^{(i)}}^{2}} PC^{(i)}   \label{PCAportfolio}
\end{equation}
where the sum is typically truncated to a small number of principal components.
  
The optimization (\ref{kellymin}) is easily generalised if we are penalising for kurtosis instead of variance; the equivalent of   (\ref{kellymin}) is 
\begin{equation}
{\bf w} = \argmax{\bf w} \left( {\bf w}.{\bf m} - \nu\ {\bf w}.{\bf w}.{\bf K}.{\bf w}.{\bf w} \right), \label{kurtmin}
\end{equation}
where $\nu$ is a different risk aversion parameter, aversion to kurtosis risk. This leads to portfolio weights
\begin{equation}
 w_{ICA}^{(i)} \propto \left( \frac{ \mu_{ICA}^{(i)}}{\kappa_{ICA}^{(i)}} \right)^{1/3}  \label{kellyICA}
\end{equation}
and the optimal portfolio is
\begin{equation}
 \Pi \propto \sum_{i} w_{ICA}^{(i)} IC^{(i)} =  \sum_{i} \left( \frac{ \mu_{ICA}^{(i)}}{\kappa_{ICA}^{(i)}} \right)^{1/3}  IC^{(i)}, \label{ICAportfolio}
\end{equation}
which can again practically be truncated to a small number of independent components. We call the portfolio (\ref{ICAportfolio}) the {\it Fat-Tailed Portfolio}.

The Fat-tailed portfolio (\ref{ICAportfolio}) has several interesting properties; in particular, the portfolio weight no longer scales linearly with the performance of a component, as it does in a  Kelly portfolio, but as its cubed root. Practically, this leads to more diversified portfolios with a less pronounced winner-takes-all profile than that dictated by Kelly. 

In particular, let us imagine two components with same variances and kurtoses, but the first one has twice the return of the second one; Kelly dictates that the first component would get twice the leverage of the second, while the fat tailed portfolio would only allocate it $2^{1/3} \approx 125.9\%$ of the leverage of the second component.

It is, however, important to note that the fat-tailed portfolio (\ref{ICAportfolio}) only maximizes the return per unit of kurtosis; it does not specifically target low variance, and its variance is not guaranteed to be low. On the other hand, the cubed root scaling serves to flatten the portfolio weights, so the portfolio is more diversified than the Kelly portfolio, as will be shown in Sections 4 and 5. 

In the absence of such diversification, the fat-tailed portfolio only moves the risk from the fourth moment to the second; it pushes the risks from the tails of the distribution towards its centre, but it does not eliminate them from the centre.

\section{Hybrid Portfolios}

We now turn to the construction of portfolios which manage both variance and kurtosis.

In analogy to (\ref{kellymin}) and (\ref{kurtmin}), one could naively look for portfolio weights by solving the combined optimization problem
\begin{equation}
{\bf w} = \argmax{\bf w} \left( {\bf w}.{\bf m}  - \lambda\ {\bf w}.{\bf V}.{\bf w} - \nu\ {\bf w}.{\bf w}.{\bf K}.{\bf w}.{\bf w} \right), \label{bothmin}
\end{equation}
where we now have two risk aversion parameters; aversion to variance, $\lambda$, and aversion to kurtosis, $\nu$. 

This is, however, not ideal. The Kelly portfolio (\ref{PCAportfolio}) and the Fat-tailed portfolio (\ref{ICAportfolio}) have a certain measure of universality; the shape of the portfolio is fixed, and the investor risk aversion only affects the leverage. With (\ref{bothmin}), this would no longer be the case. The shape of the portfolio would now depend on the ratio of variance aversion to kurtosis aversion, and it woud be different from investor to investor.

We are, on the other hand, interested in finding universal portfolio shapes that control both variance and kurtosis, while being independent of the investor risk preferences. In other words, the investor risk preferences should only affect the  leverage, but not the shape of the portfolio.

The solution comes from the Central Limit Theorem, which loosely states that the sum of independent random variables, de-meaned and normalized to unit volatility, tends to a normal distribution. In particular, its excess kurtosis vanishes due to the Central Limit Theorem, and its variance decays through diversification.

We therefore look at portfolios of ICs that satisfy the conditions of the Central Limit Theorem. For this, we have the following result:

\begin{theorem}[Central Limit Theorem for ICA]
Consider two portfolios,
$$ \Pi_{PC} = \frac{1}{n}   \sum_{i}^{n}  \frac{   PC^{(i)} }{ \sigma_{PCA}^{(i)}} ,$$
$$ \Pi_{IC} = \frac{1}{n}    \sum_{i}^{n}  \frac{   IC^{(i)} }{ \sigma_{ICA}^{(i)}}.$$
Then, as $n\rightarrow \infty$, the variance of the returns of both portfolios is 
$$O\left(\frac{1}{n}\right);$$
the excess kurtosis of the returns of $ \Pi_{PC}$ is
$$o\left(\frac{1}{n}\right),$$
and the excess kurtosis of the returns of $ \Pi_{IC}$ is
$$o\left(\frac{1}{n^{2}}\right).$$
\end{theorem}

The proof of the CLT for ICA is given in the Appendix. Note that there is formally no need to normalise the components of $ \Pi_{IC}$ by $\sigma_{ICA}^{(i)}$ in constructing the IC portfolio, as we do in the statement of the CLT, since volatilities of ICs are already all equal by construction. We are only normalising them so that we would be able to compare them directly to PCA portfolios in the numerical example in the next Section.

This form of the CLT tells us that portfolios of ICs can be equally good as portfolios of PCs in suppressing the variance, but they have the added bonus of suppressing the kurtosis of the returns much faster.

So, a portfolio of 10 PCs would reduce the excess kurtosis by 90\%, but a portfolio of 10 ICs would reduce it by 99\%; to obtain the equivalent 99\% reduction using PCs, we would need to sum 100 PCs.

To estimate how serious this difference is, we turn to a bit of dimensional analysis to compare these quantities in different portfolios.

If we view the kurtosis of a process as the variance of its square, modulo constant, we get the usual interpretation of kurtosis as variance-of-variance. In other words, the excess kurtosis is roughly proportional to the volatility-of-volatility to the power of 4. Therefore, the CLT tells us that, in the worst case scenario, PCA portfolios have volatility decaying as $1/\sqrt{n}$, and volatility-of-volatility decaying as $n^{-1/4}$; while ICA portfolios have both volatility and volatility-of-volatility decaying as $n^{-1/2}$.

In other words, PCA portfolios can, in the worst case scenario, become more leptokurtic as the number of components increases; the ratio of volatility-of-volatility to volatility grows like $n^{1/4}$. For ICA portfolios, the ratio of volatility-of-volatility to volatility is stationary, and the worst case scenario is that the leptokurticity remains constant.

\section{S\&P500 stocks}

We looked at S\&P500 stocks over a period of 12 years, from the 1st January 2007 until the 31st December 2018. To counteract the effects of stocks drifting in and out of the index over such a long time frame, we have divided the time frame into four buckets, each lasting three calendar years; from 1st January 2007 until 31st December 2009, from 1st January 2010 until 31st December 2012, from 1st January 2013 until 31st December 2015 and from 1st January 2016 until 31st December 2018. The basket for each bucket was selected as consisting of the index constituents on the last business day prior to the start of the bucket, and these stocks were followed until the end of the bucket. Any stock that was de-listed before the end of a bucket in which it appeared was deemed to have returned $0\%$ from its last trading day until the end of the bucket. There were no adjustments for stocks entering or leaving the index over the duration of any of the buckets.

We have performed PCA and ICA on each of the buckets, extracting 10 principal components and 10 independent components for each bucket. Decompositions used the Python package \texttt{scikit-learn 0.23.2}, and in particuar the classes  \texttt{sklearn.decomposition.PCA} for PCA, and \texttt{sklearn.decomposition.fastICA} for ICA.

\begin{figure}[p]
\centering
\includegraphics[width=6cm]{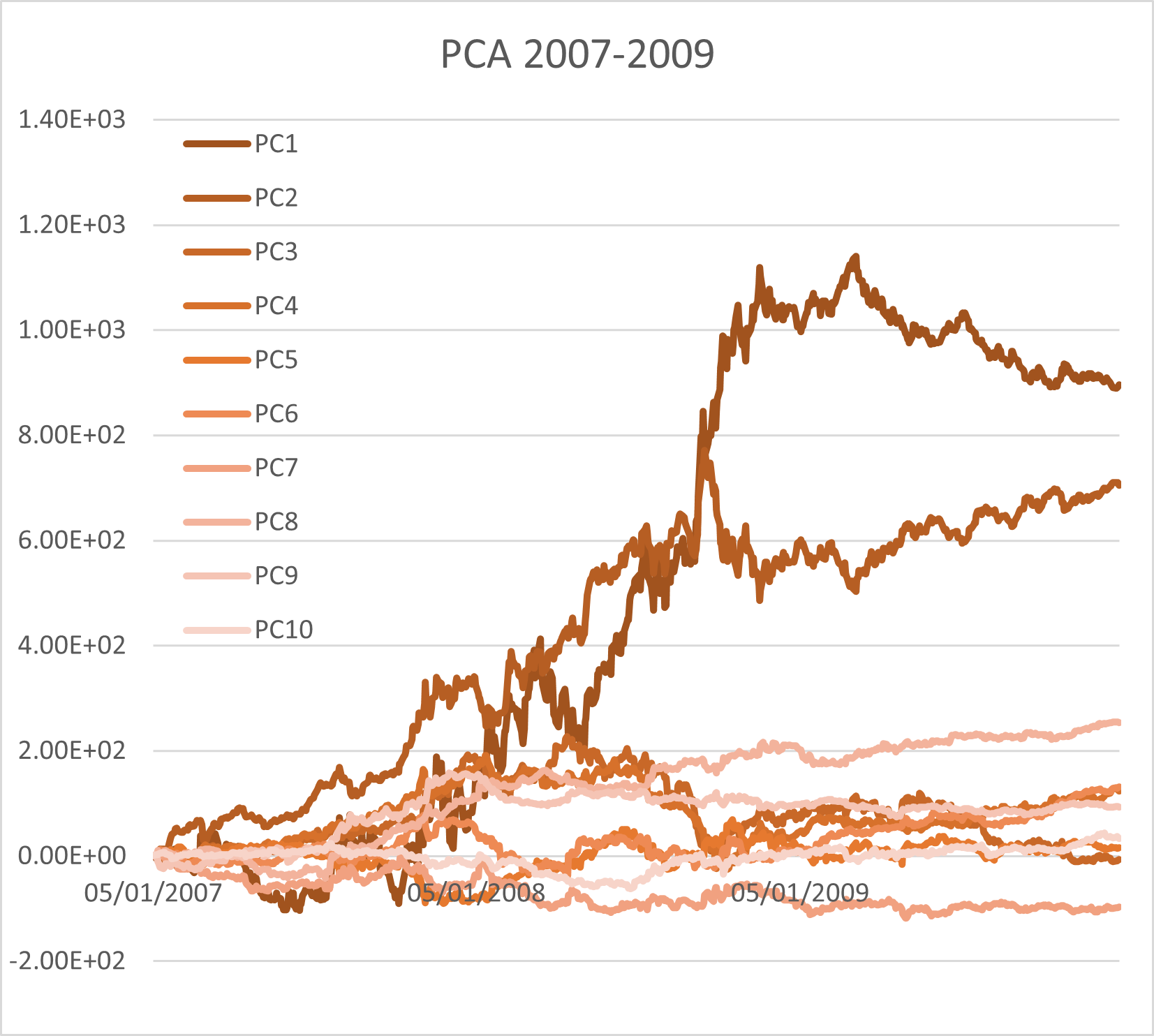}
\includegraphics[width=6cm]{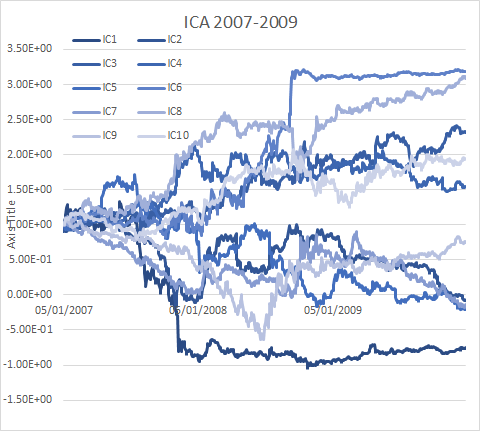}
\includegraphics[width=6cm]{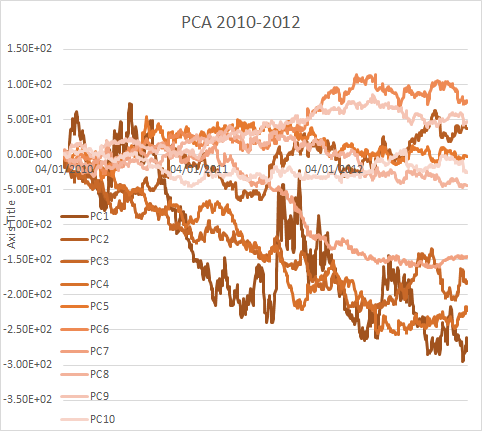}
\includegraphics[width=6cm]{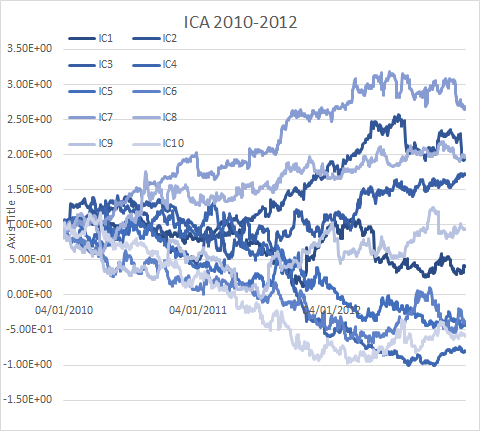}
\includegraphics[width=6cm]{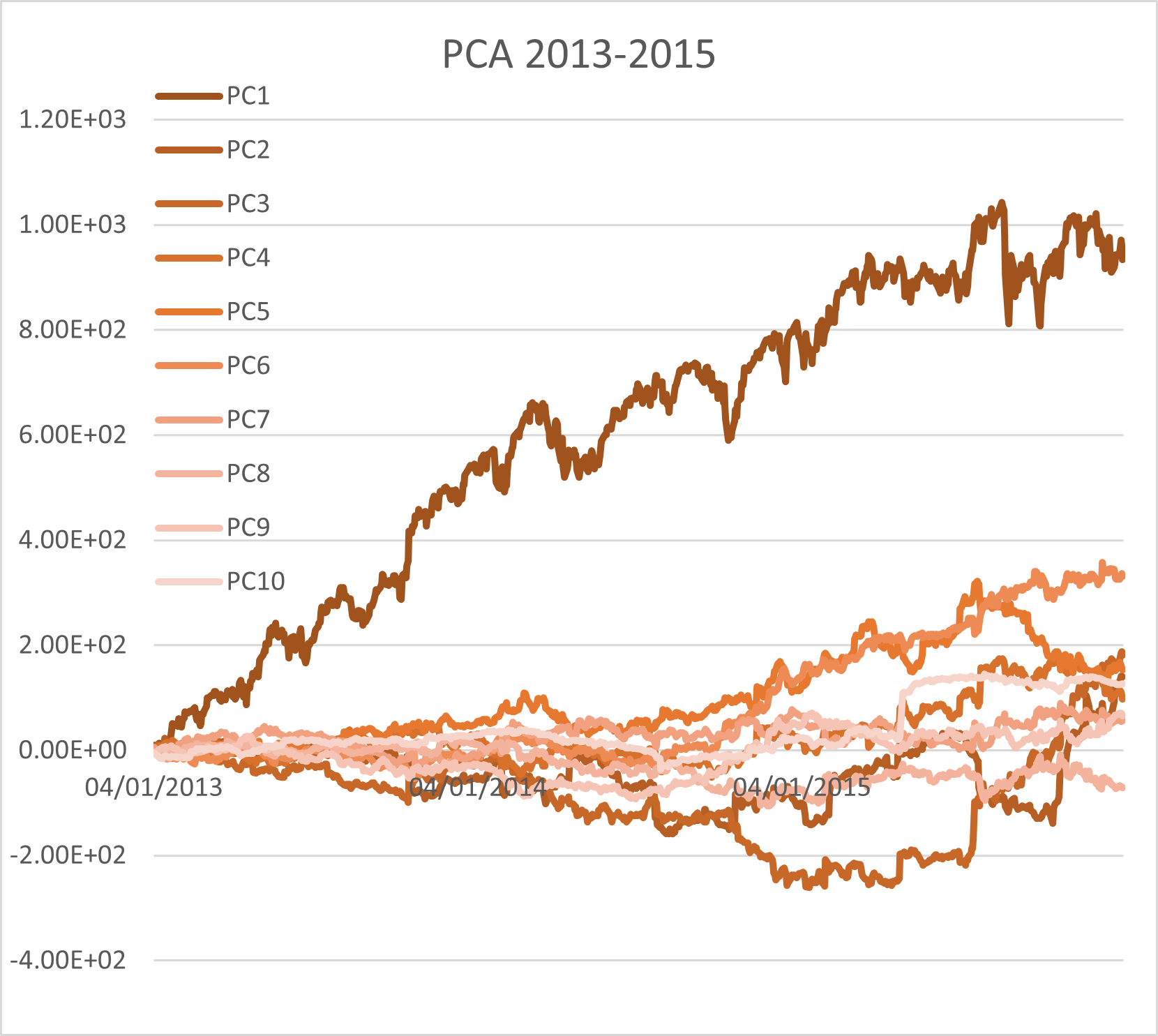}
\includegraphics[width=6cm]{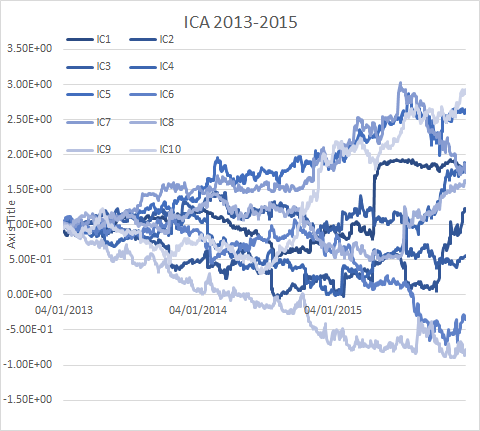}
\includegraphics[width=6cm]{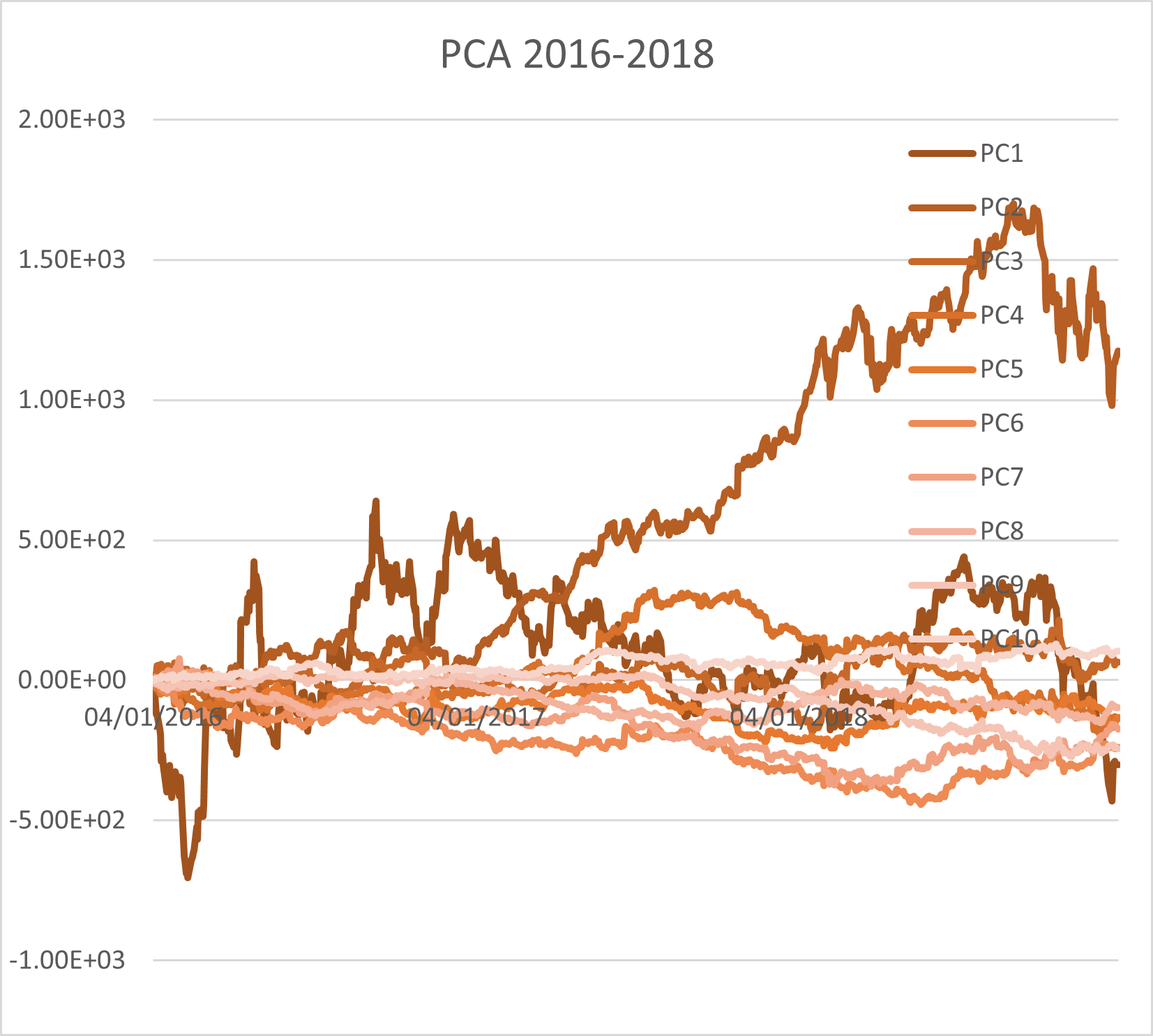}
\includegraphics[width=6cm]{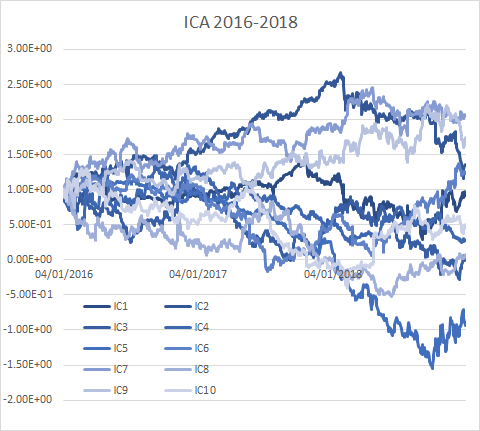}
\centering
\caption{First ten PCs and first ten ICs,  for each of the 2007-2009, 2010-2012, 2013-2015 and 2016-2018 buckets.}
\label{fig:PCsICs}
\end{figure}

Resulting PCs and ICs are plotted in Figure \ref{fig:PCsICs}. Qualitatively, the PCs appear to be significantly more dispersed than the ICs; PC1 is immediately visually identifiable as an outlier on each PCA graph, while IC1 is not an obvious outlier in any of the buckets.

\begin{figure}[p]
\centering
\includegraphics[width=10cm]{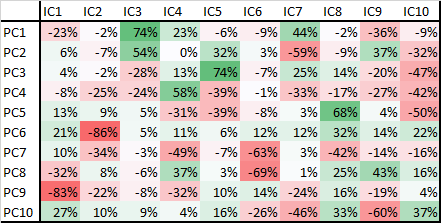}
\includegraphics[width=10cm]{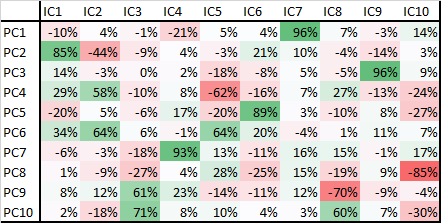}
\includegraphics[width=10cm]{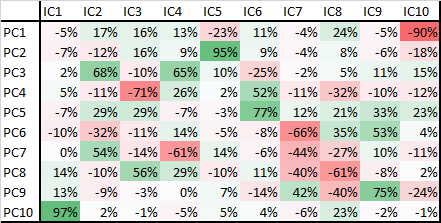}
\includegraphics[width=10cm]{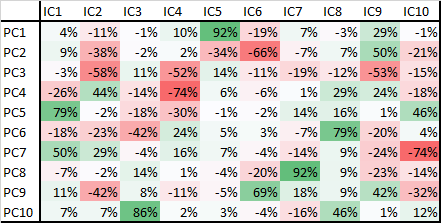}
\centering
\caption{Correlations between the PCs and ICs for the 2007-2009 bucket (top), 2010-2012 (second from the top), 2013-2015 (second from the bottom) and 2016-2019 (bottom) buckets.}
\label{fig:correl}
\end{figure}

The correlation matrices between the PCs and the ICs in each bucket are shown in Figure \ref{fig:correl}. PCs are generally not orthogonal to the ICs, and the correlation matrix has a block structure, with the PC-IC block generally non-zero. This is significant insofar as it illustrates that ICs are not just PCs by another name. Even though the top PCs and the top ICs span roughly the same space, they are not the same and the transformation from PCs to ICs is non-trivial. By extension, this is also a test of non-Gaussianity; if the underlying processes were generated by a multivariate Gaussian variable, the PCs and the ICs would coincide exactly; kurtosis of a Gaussian variable is $\propto \sigma^{4}$, so a component selected for the highest variance (PC)  would also be selected for the highest kurtosis (IC). This is clearly not the case for S\&P500 stocks.

\begin{figure}[p]
\centering
\includegraphics[width=7cm]{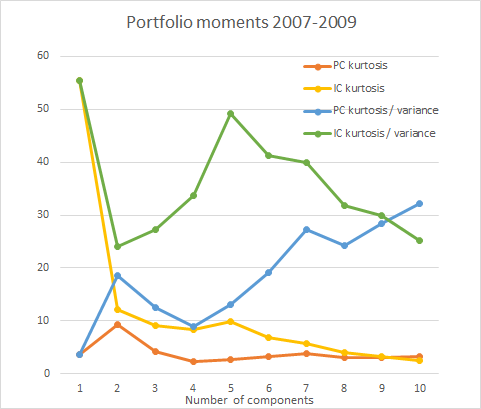}
\centering
\includegraphics[width=7cm]{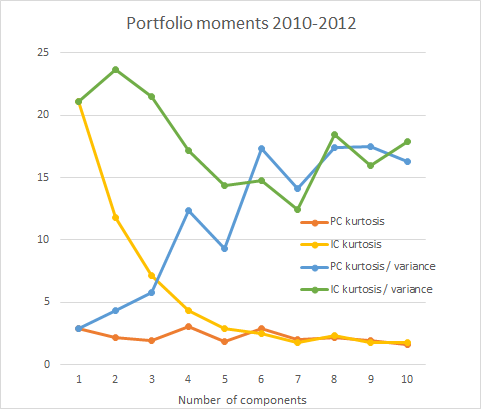}
\centering
\includegraphics[width=7cm]{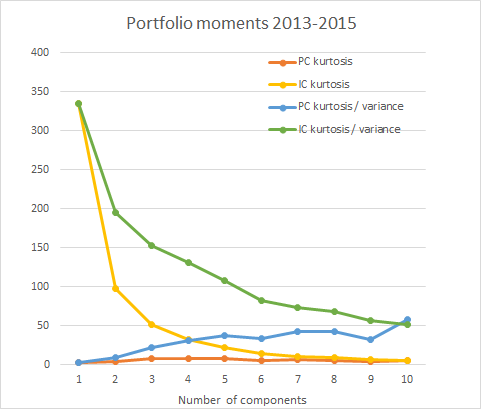}
\centering
\includegraphics[width=7cm]{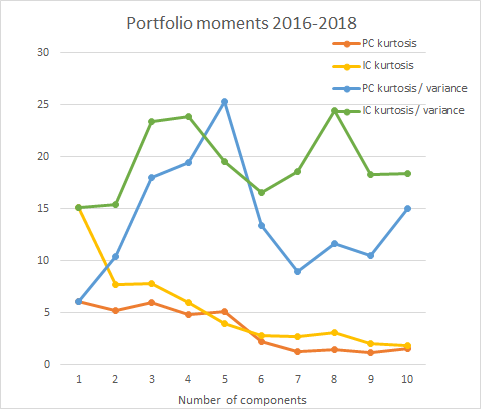}
\centering
\caption{Kurtosis and kurtosis/variance ratio of equal volatility baskets with increasing portfolio size $n$, PC portfolios vs IC portfolios. }
\label{fig:moments}
\end{figure}

We have constructed the portfolios $ \Pi_{PC}$ and $ \Pi_{IC}$ from the Central Limit Theorem from the increasing number of components $n$, starting from a single component and ending with all $n=10$ components. The variances and kurtoses of the resulting portfolios are shown in Figure \ref{fig:moments}. Each component was scaled to unit variance, so the variance of each portfolo is exactly $1/n$, to machine prexision.  The kurtosis illustrates the difference between the portfolios; kurtosis of  $ \Pi_{IC}$ decays faster than the variance with increasing $n$, while the kurtosis of $ \Pi_{PC}$ generally does not. This is  consistent with the respective $n^{-1}$ and $n^{-2}$ scaling for the kurtoses of $\Pi_{PC}$ and $\Pi_{IC}$ as predicted by the Central Limit Theorem. 

\begin{figure}[p]
\centering
\includegraphics[width=7cm]{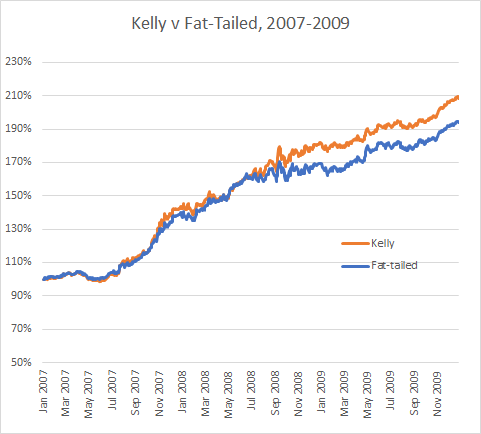}
\centering
\includegraphics[width=7cm]{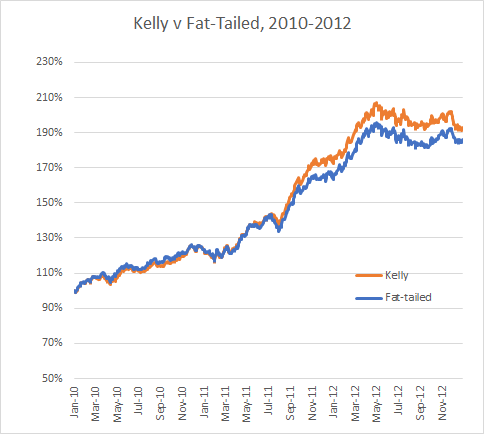}
\centering
\includegraphics[width=7cm]{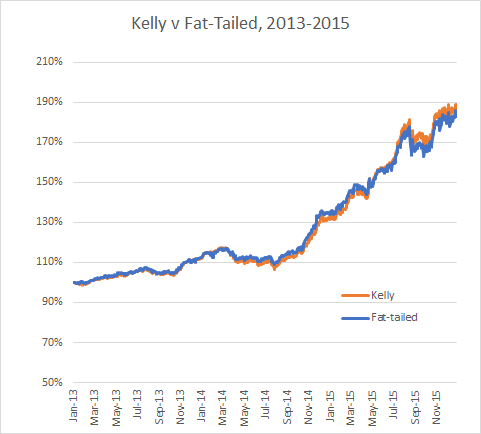}
\centering
\includegraphics[width=7cm]{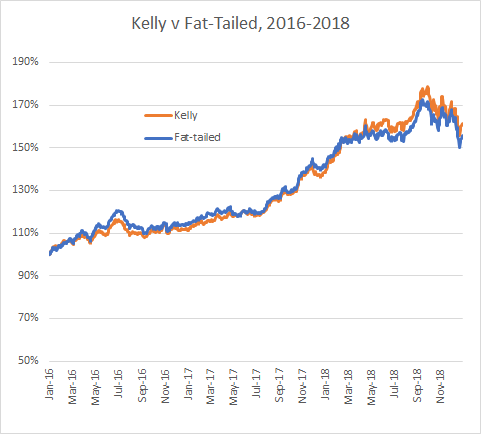}
\centering
\includegraphics[width=15cm]{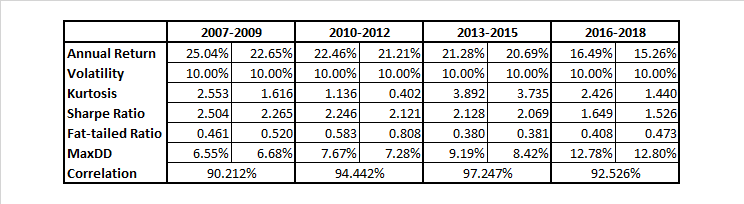}
\caption{Kelly portfolio vs Fat-tailed portfolio for each bucket; Fat-tailed Ratio refers to the quantity $\left( \mu / \kappa \right)^{1/3}$, and MaxDD is the maximum drawdown}.
\label{fig:portfolio}
\end{figure}

Finally, we have constructed the optimal Kelly portfolio from the ten PCs, and the optimal Fat-tailed portfolio from the ten ICs, scaled to the same portfolio volatility of $10\%$ in all cases. The resulting portfolios are shown in Figure  \ref{fig:portfolio}. The correlation between the Kelly portfolio and the Fat-tailed portfolio is greater than $90\%$ for each bucket, and it is as high as $97.247\%$ in the 2013-2015 bucket. Given that the total number of components is 10, correlation over $90\%$ implies that both Kelly and Fat-tailed portfolios capture the same factors and the same performance over each bucket. This is indeed the case in our portfolios.

Unsurprisingly, due to their construction, the Kelly portfolio always has lower volatility and higher Sharpe ratio, while the Fat-tailed portfolio always has lower kurtosis and higher Fat-tailed ratio $\left( \mu / \kappa \right)^{1/3}$. The differences between the Sharpe ratios of the Kelly and Fat-tailed portfolios are always below $10\%$. Given such small differences for {\it a posteriori} portfolio construction with perfect hindsight, in real-world, {\it a priori} portfolio construction without perfect hindsight, the differences in Sharpe ratios are unlikely to be observable to some reasonable level of confidence. 

As a guideline, each of our 3-year buckets consists of approximately 756 trading days, implying the error in the estimation of mean as $1/\sqrt{756} \approx 3.65\%$. Therefore, our Sharpe ratios can at best be correct to $\pm 3.65\%$ {\it a posteriori}. Adding a similar uncertainty for the estimation of volatility and moving into the {\it a proiri} world with uncertainty over parameter drift, the acuracy of our Sharpe ratio estimates is unlikely to be better than $\pm 10\%$. Hence, any differences in Sharpe ratios of the order of 10\% or less are likely immaterial.

The differences between the kurtoses and Fat-tailed ratios between the two portfolio classes vary more widely, ranging from $10\%$ difference in 2013-2015 (0.49 vs 0.54) to $60\%$ difference in 2010-2012 (0.766 vs 1.226). Some of these differences would arguably survive into practical {\it a priori} portfolio construction, where they would be felt.

We can see an indirect confirmation of this by looking at the maximum drawdown of the selected portfolios, as an independent benchmark of the success of the two portfolio construction methods. The  Kelly portfolio has smaller  maximum drawdown than the Fat-tailed portfoilio in the 2007-2009 bucket (6.5\% vs 6.7\%), larger in the 2010-2012 and 2013-2015 buckets (7.7\% vs 7.3\% and 9.2\% vs 8.4\%), and the drawdowns are indistinguishable in the 2016-2018 bucket (12.8\% vs 12.8\%). While far from conclusive, this supports the notion that fat-tailed portfolio construction keeps a better handle on the overall portfolio risks per unit of volatility.

\section{Discussion}

In the context of optimal portfolio construction in the presence of fat tails, non-Gaussianity-based factors such as those described in this paper are an interesting alternative to standard PCA-based factors. 

While PCA and ICA are conceptually similar orthogonalization methods, their stated purpose is different. The purpose of PCA is to isolate the strongest signals (those with highest variance), while the purpose of ICA is to  isolate the noisiest signals (those with highest deviation from Gaussianity). 

This leads to a different distribution of factors, as illustrated in our S\&P500 example. The first IC is nowhere near as dominant as the first PC, which directly leads to the slower, $1/3$ scaling in the fat-tailed portfolio construction. This, in turn, reduces portfolio concentration and generates smoother, more diversified portfolios. As illustrated in our example, fat-tailed portfolios end up capturing the same factors as Kelly portfolios, but they manage the fat-tailed risks better due to their $1/3$ scaling, as opposed to linear scaling of the Kelly portfolios.

Despite this lesser dispersion of ICs when compared to PCs, the Central Limit Theorem for ICs shows that they are nonetheless good diversifiers; in many respects, they are better diversifiers than PCs, in the sense that they diversify variance equally well, but they also diversify kurtosis much faster. The numerical results in Figure \ref{fig:moments} confirm this in our specific example.

There are two main advantages of our Fat-taied portfolio construction over Kelly portfolios; one based on the properties of the ICA as opposed to the PCA, and the other based on the $1/3$ scaling of the weights.

The key advantage of the ICA is that makes no assumptions about Gaussianity of the underlying porcesses, and hence it does not automatically break down when such assumptions are not met. We can therefore expect ICA-based components to be more robust than PCA-based components.

The key advantage of the $1/3$ scaling of the portfolio weight is that it is sub-linear, and that the portfolio weight changes more slowly with increasing $\mu$ and decreasing $\kappa$. Therefore, errors in the estimation of these quantities will have a smaller effect on portfolio weights, and tus on the overall shape of the portfolio.

ICA does not solve all known problems of PCA. The problem of low confidence in asset correlations is not fixed by moving to higher order codependence; it arguably becomes worse, due to the same number of data points being used to infer a higher-order statistic. On the other hand, as addresed in the previous paragraph, the $1/3$ scaling of portfolio weights means that the portfolio construction is  less sensitive to the exact estimation of codependence numbers.

\section{Conclusions}

We have proposed a new, ICA-based Fat-tailed portfolio construction method. The method rests on identifying the statistically independent components through ICA, and then weighting such components in proportion to the ratio of their return to kurtosis, raised to the power of $1/3$.

This method is shown to have the following advantages over covariance-based Kelly portfolios:
\begin{itemize}
\item ICA is more robust than PCA in the face of non-Gaussianity of the underlying processes
\item the $1/3$ portfolio scaling is more robust than the linear scaling in the face of uncertainties in parameter estimation
\item the resulting portfolios maintain better control of tail risks
\end{itemize}

These advantages do not come at a cost in terms of increased computational or operational complexity; ICA calculation is available  as part of numerous publicly available numerical packages, and the only additional calculation  is the calculation of the kurtosis of each component, which is equally  straightforward.

In conclusion, ICA based factor analysis and portfolio construction methods are a powerful alternative to the existing array of methods, and they address a number of known concerns with covariance-based methods.

\section*{Appendix: Proof of the Central Limit Theorem for ICA}

The CLT for $\Pi_{PC}$ is the standard CLT for any sum of de-meaned, normalized orthogonal random variables.
We denote 
\begin{equation}
Y_{i} = \frac{  d  PC^{(i)} - \mu_{PCA}^{(i)}}{ \sigma_{PCA}^{(i)}} \label{Y}
\end{equation}
where $d  PC^{(i)}$ denotes the returns process of $ PC^{(i)}$, so that all $Y_{i}$ all have zero mean and unit variance, and their normalised sums as
\begin{equation}
Z_{n} = \frac{1}{\sqrt{n}} \sum_{i}^{n} Y_{i} \label{Z}
\end{equation}

Given that $Y_{i}$ are all orthogonal, the characteristic function of $Z_{n}$ is
\begin{align}
\varphi_{Z_{n}} (t) &= \varphi_{\frac{1}{\sqrt{n}} \sum_{i}^{n} Y_{i}} (t)  \\
&= \varphi_{Y_{1}} \left( \frac{t}{\sqrt{n}} \right) \varphi_{Y_{2}} \left( \frac{t}{\sqrt{n}} \right) ...\ \varphi_{Y_{n}} \left( \frac{t}{\sqrt{n}} \right) + o\left(\frac{t^{2}}{n} \right) \label{eq13}\\
&=\left[ \varphi_{Y_{1}} \left( \frac{t}{\sqrt{n}} \right)  \right]^{n}+ o\left(\frac{t^{2}}{n} \right)
\end{align}
as $t^{2}/n \rightarrow 0$.

The error term $o(t^{2}/n)$ arises in equation (\ref{eq13}) due to the fact that we can not guarantee that joint moments beyond order 2 are zero. The construction of PCs only guarantees their orthogonality, but it does  not guarantee either their Gaussianity, nor their independence at orders higher than 2.

The second order Taylor expansion of $\varphi_{Y_{1}}$ around zero gives
\begin{equation}
\varphi_{Y_{1}} = \left( 1 - \frac{t^{2}}{2n} + o\left(\frac{t^{2}}{n}  \right) \right),
\label{taylor}
\end{equation}
so 
\begin{equation}
\varphi_{Z_{n}} = \left( 1 - \frac{t^{2}}{2n} + o\left(\frac{t^{2}}{n}  \right) \right)^{n}  + o\left(\frac{t^{2}}{n} \right) .
\label{product}
\end{equation}
Expanding the known terms in (\ref{product}) to fourth order, we get
\begin{equation}
\varphi_{Z_{n}} =  1 - n \frac{t^{2}}{2n} + \frac{n(n-1)}{2}  \frac{t^{4}}{4n^{2}}+ o\left(\frac{t^{2}}{n} \right)
\label{expansion}
\end{equation}
where the term $o(t^{2}/n)$ may contain additional contributions at the order $t^{4}$, so the best we can say about the error at $t^{4}$ is that it is $o(1/n)$.

In particular, the second and fourth derivatives of $\varphi_{Z_{n}}$ at zero are
\begin{align}
\varphi_{Z_{n}}'' &= 1   \\
\varphi_{Z_{n}}^{iv} &=   \frac{n(n-1)}{n^{2}} 3  +o\left( \frac{1}{n} \right),
\end{align}
hence the excess kurtosis of $Z_{n}$ is $o(1/n)$. Given that the returns of $\Pi_{PC}$ differ from $Z_{n}/\sqrt{n}$ by a deterministic mean process, the excess kurtosis of the returns of $\Pi_{PC}$ is also $o(1/n)$.

Moving to $\Pi_{IC}$, the difference is that the components are now, by construction,  independent to the fourth order, instead of to second order for $\Pi_{PC}$. Therefore, denoting 
\begin{equation}
y_{i} = \frac{ d  IC^{(i)} - \mu_{ICA}^{(i)}}{ \sigma_{ICA}^{(i)}}, \label{yy}
\end{equation}
where $d IC^{(i)}$ denotes the returns process of  $ IC^{(i)}$
\begin{equation}
z_{n} = \frac{1}{\sqrt{n}} \sum_{i}^{n} y_{i}, \label{zz}
\end{equation}
 equations (\ref{taylor}) and (\ref{product}) read 
\begin{equation}
\varphi_{y_{1}} = \left( 1 - \frac{t^{2}}{2n} + o\left(\frac{t^{4}}{n^{2}}  \right) \right),
\label{taylor1}
\end{equation}
\begin{equation}
\varphi_{z_{n}} = \left( 1 - \frac{t^{2}}{2n} + o\left(\frac{t^{4}}{n^{2}} \right)  \right)^{n}  + o\left(\frac{t^{4}}{n^{2}} \right),
\label{product1}
\end{equation}
with the error at $t^{4}$ now of the order $o(1/n^{2})$.

The second and fourth derivatives become
\begin{align}
\varphi_{Z_{n}}'' &= 1   \\
\varphi_{Z_{n}}^{iv} &=   \frac{n(n-1)}{n^{2}} 3  +o\left( \frac{1}{n^{2}} \right),
\end{align}
and the excess kurtoses of $z_{n}$ and the returns of  $\Pi_{IC}$ are  $o(1/n^{2})$. 

\section*{Acknowledgments}

The author reports no conflicts of interest. The author alone is responsible for the content and writing of the paper.

\begin{thebibliography}{1}


\bibitem{avellaneda} Avellaneda, M.   (2019) \emph{Hierarchical PCA and Applications to Portfolio Management}, {https://ssrn.com/abstract=3467712 or http://dx.doi.org/10.2139/ssrn.3467712 }
\bibitem{principal}  Avellaneda, M., Healy, B.,  Papanicolaou,  A. \& Papanicolaou, G. (2020) \emph{Principal Eigenportfolios for US Equities}. https://papers.ssrn.com/sol3/papers.cfm?abstract\_id=3738769

\bibitem{choi} Choi, S., Cihocki, A., Park, H.-M., \& Lee, S.-Y. (2005) \emph{Blind source separation and independent component analysis: A review.}  Neural Information Processing-Letters and Reviews 6.1: 1-57.
\bibitem{audio} Haykin, S \&  Kan, K. (2007) \emph{Coherent ICA: Implications for Auditory Signal Processing}, 1 - 5. 10.1109/ASPAA.2007.4393059. 
\bibitem{hyva} Hyv\"{a}rinen, A.  (2013) {Independent component analysis: recent advances}, \emph{Philosophical Transactions: Mathematical, Physical and Engineering Sciences}. 371
\bibitem{pca} Jollife, I.T. (2002) \emph{Principal  Compoment  Analysis},  2nd  edition,  Springer,  NewYork.

\bibitem{mark1} Markowitz, H.M. (1952) {Portfolio Selection}, \emph{The Journal of Finance}. 7 (1): 77–91. doi:10.2307/2975974. JSTOR 2975974.
\bibitem{mark2}  Markowitz, H.M. (1956) {The Optimization of a Quadratic Function Subject to Linear Constraints}, \emph{Naval Research Logistics Quarterly}. 3 (1–2): 111–133. doi:10.1002/nav.3800030110.
\bibitem{visual} Martín-Clemente, R. \& Hornillo-Mellado, S. (2006) \emph{Image processing using ICA: a new perspective}, IEEE MELECON 2006, May 16-19, Benalmádena (Málaga), Spain.
\bibitem{telecom}  Parmar, S. D. \& Unhelkar,  B. (2009) {Separation performance of ICA algorithms in communication systems},  \emph{International Multimedia, Signal Processing and Communication Technologies}, Aligarh, pp. 142-145, doi: 10.1109/MSPCT.2009.5164195.
\bibitem{shkolnik}  Shkolnik, A.D., Goldberg, L. \& Bohn, J.R.   (2016) \emph{   Identifying   broad   andnarrow    financial    risk    factors    with    convex    optimization},  https://ssrn.com/abstract=2800237  or http://dx.doi.org/10.2139/ssrn.2800237
\bibitem{eeg}Ungureanu, M. , Bigan, C., Strungaru,  R.  \& Lazarescu, V.  (2004) \emph{Independent Component Analysis Applied in Biomedical Signal Processing}, Meas Sci Rev, Volume 4, Section 2.
\bibitem{icawiki} Hyv\"{a}rinen A., Karhunen, J. \& Oja, E. (2001) \emph{Independent component analysis}, John Wiley \& Sons, DOI:10.1002/0471221317



\end{thebibliography}
\end{document}